\documentclass[aps,prl,twocolumn,epsfig,preprintnumbers,groupedaddress,10pt]{revtex4}

\usepackage{graphicx}
\usepackage{dcolumn}
\usepackage{bm}

\newcommand{\be}{\begin{equation}}
\newcommand{\ee}{\end{equation}}
\newcommand{\bea}{\begin{eqnarray}}
\newcommand{\eea}{\end{eqnarray}}
\newcommand{\ba}{\begin{eqnarray}}
\newcommand{\ea}{\end{eqnarray}}

\newcommand{\beq}{\begin{equation}}
\newcommand{\eeq}{\end{equation}}
\newcommand{\beqa}{\begin{eqnarray}}
\newcommand{\eeqa}{\end{eqnarray}}
\newcommand{\beqar}{\begin{eqnarray*}}
\newcommand{\eeqar}{\end{eqnarray*}}

\begin{document}

\preprint{CERN-PH-TH/2010-028}

\title{Testing nuclear parton distributions with pA collisions at the LHC}
\author{Paloma Quiroga-Arias$^{1,3}$, Jos\'e Guilherme Milhano$^{2,3}$ and Urs Achin Wiedemann$^{3}$}

\address{$^1$ Departamento de F\'isica de Part\'iculas and IGFAE, Universidade de Santiago de Compostela 15706 Santiago de Compostela, Spain\\
$^2$ CENTRA, Departamento de F\'isica, Instituto Superior T\'ecnico (IST),
Av. Rovisco Pais 1, P-1049-001 Lisboa, Portugal\\
$^3$ Physics Department, 
    Theory Unit, CERN,
    CH-1211 Gen\`eve 23, Switzerland}


\begin{abstract}
Global perturbative QCD analyses, based on large data sets from electron-proton and 
hadron collider experiments, provide tight constraints on the parton distribution function (PDF)
in the proton. The extension of these analyses to nuclear parton distributions (nPDF) has attracted
much interest in recent years. nPDFs are needed as benchmarks for the characterization of
hot QCD matter in nucleus-nucleus collisions, and attract further interest since they may show
novel signatures of non-linear density-dependent QCD evolution. However, it is not known
from first principles
whether the factorization of long-range phenomena into process-independent parton distribution,
which underlies global PDF extractions for the proton, extends to nuclear effects. As a consequence,
assessing the reliability of nPDFs for benchmark calculations goes beyond testing the 
numerical accuracy of their extraction and requires phenomenological tests of the factorization assumption. Here we argue that a proton-nucleus collision program at the LHC would provide
a set of measurements allowing for unprecedented tests of the factorization assumption
underlying global nPDF fits.
\end{abstract}

\maketitle


Parton distribution functions (PDFs) $f_{i/h}(x,Q^2)$ play a central role in the study of
high energy collisions involving hadronic projectiles $h$. They define the
flux of quarks and gluons ($i = q, g$) in hadrons as a function of the partonic resolution scale
$Q^2$ and hadronic momentum fraction $x$. For protons, sets of collinearly factorized
universal PDFs have been obtained, since a long time ago, in global perturbative QCD analyses. 
These are 
based on data from deep-inelastic lepton-proton scattering (DIS) and Drell-Yan (DY) production, 
as well as W/Z and jet production at hadron colliders. These data provide tight constraints
on PDFs over logarithmically wide ranges in $Q^2$ and $x$, and have allowed precision 
testing of linear perturbative QCD evolution.
In comparison to proton PDFs, our understanding of parton distribution functions $f_{i/A}(x,Q^2)$ in
nuclei of nucleon number $A$ is much less mature. Knowledge of nuclear parton distribution functions
(nPDFs) is important  in heavy ion collisions at RHIC and at the LHC for a quantitative control
of hard processes, which are employed as probes of dense QCD matter.
Characterizing nuclear modifications of PDFs is also of great interest in its own right, 
since the nuclear environment is expected to enhance parton density-dependent effects, which
can reveal qualitatively novel, non-linear features of QCD evolution.
 
Paralleling the determination of proton PDFs, several global
QCD analyses of nPDFs have been made within the last decade~\cite{Eskola:2009uj,Eskola:2008ca,Eskola:1998df,deFlorian:2003qf,Hirai:2007sx}.
Up until recently, these analyses were based solely on fixed-target nuclear DIS and DY data.
Compared to the data constraining proton PDFs, these are 
of lower precision and lie in a much more limited range of $Q^2$ and $x$. 
Constraints on nuclear gluon distribution functions are particularly poor, since they cannot
be obtained from the absolute values of DIS structure functions,  but only from their 
logarithmic $Q^2$-evolution, for which a wide $Q^2$-range is mandatory. 
To improve on this deficiency, recent global nPDF analysis~\cite{Eskola:2008ca,Eskola:2009uj} have included for the first time 
data from inclusive high-$p_T$ hadron production in hadron-nucleus scattering measured 
at RHIC~\cite{Adler:2006wg,Adams:2006nd,Arsene:2004ux}. 

However, in contrast to the theoretical basis for global analyses of proton PDFs, 
the separability of nuclear effects into process-independent nPDFs and process-dependent 
but A-independent hard processes is not established within the framework of collinear 
factorized QCD. In particular, some of the characteristic nuclear dependencies in 
hadron-nucleus collisions, such as the Cronin effect~\cite{Cronin:zm}, may have a 
dynamical origin that cannot, or can only partly, be absorbed in 
process-independent nPDFs. In view of the importance of nPDFs for characterizing 
benchmark processes in heavy ion collisions, it is thus desirable to look for stringent
phenomenological tests of the working assumption of global nPDF fits that 
the dominant nuclear effects can be factorized into the incoming PDFs.
Here, we argue that a program of hadron-nucleus collisions at the LHC would provide for such 
tests with unprecedented quality. 

\begin{figure}
\vskip -.7cm
\centering
\includegraphics[width=1.02\linewidth]{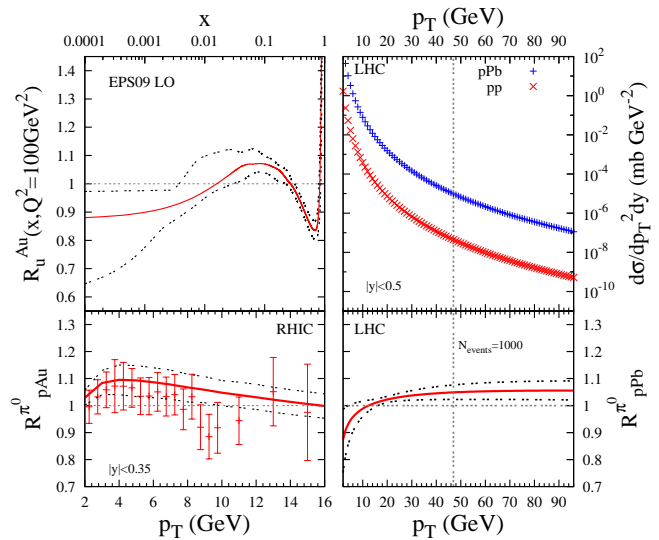}
\vskip -0.7 cm
\caption{\label{fig1} (left,top) The ratio (\ref{eq2}) of nuclear to nucleon PDFS for 
valence up-quarks at $Q^2 = (10\, {\rm GeV})^2$ obtained in the EPS09 LO analysis. 
Dashed lines characterize the range of uncertainties.  (left,bottom) The nuclear modification
factor (\ref{eq3}) for neutral pion production in $\sqrt{s_{\rm NN}} = 200$ GeV dAu (RHIC). Data from PHENIX \cite{Adler:2006wg} are compared to a EPS09 LO calculation.
Hereafter, uncertainty bands are from EPS09 LO only (uncertainties from proton PDFs and FF are neglected).
(right, bottom) ibidem, for 
$\sqrt{s_{\rm NN}} = 8.8$ TeV pPb (LHC). 
(right, top) The corresponding single inclusive
pion spectra. The thin vertical line denotes the kinematic range with statistics 
of more than 1000 events per GeV-bin after one month of LHC operation with pPb.
}
\vskip -0.5 cm
\end{figure}

We will focus mainly on single inclusive high-$p_T$ hadron production.
In the factorized QCD ansatz to hadron-nucleus collisions, the cross section for production
of a hadron $h$ takes the form 
\begin{equation}
	d^3\sigma^{pA \to h\, X} = 
	A\, \sum_{ijk} f_{j/p}\, f_{i/A}
	\otimes d^3\sigma^{ij \to k\, X} \otimes D_{k\to h}\, ,
	\label{eq1}
\end{equation}
where the symbol $\otimes$ stands for the convolution of the incoming PDFs with the
cross section of the hard partonic process and with the fragmentation function for
a parton $k$ into a hadron $h$. The sum goes over all parton species contributing to the
production of $h$. By construction, the entire nuclear dependence of the cross section 
(\ref{eq1}) resides in the nPDF $f_{i/A}(x,Q^2)$. It is customary to characterize nuclear 
effects by the ratios 
\begin{equation}
  R_i^A(x,Q^2) \equiv	f_{i/A}(x,Q^2) \big/ f_{i/p}(x,Q^2)\, .
  \label{eq2}
\end{equation}
In global nPDF analyses, characteristic deviations of $R_i^A(x,Q^2)$ from unity are found
for all scales of $Q^2$ tested so far and for essentially all scales of the momentum fraction $x$.
These effects are typically referred to as nuclear shadowing
($x \lesssim 0.01$), anti-shadowing ($0.01 \lesssim x \lesssim 0.2$), EMC effect ($0.2 \lesssim x \lesssim 0.7$) and Fermi motion ($x \gtrsim 0.7$). A 
typical example for  the nuclear $x$-dependence of
$R_i^A(x,Q^2)$ is shown in the upper left plot of Fig.~\ref{fig1}.

Nuclear effects on single inclusive hadron production are typically characterized 
by  the nuclear modification factor $R_{p\, A}^{h}$, which depends on 
the transverse momentum $p_T$ and the rapidity $y$ of the 
hadron, 
\begin{equation}
 	R_{p\, A}^{h}(p_T, y) = \frac{d\sigma^{pA \to h+X}}{dp_T^2\, dy}  \Bigg /  
	N_{\rm coll}^{pA} \frac{d\sigma^{pp \to h+X}}{dp_T^2\, dy}\, .
	\label{eq3}
\end{equation}
Here, $N_{\rm coll}^{pA}$ denotes the average number of equivalent nucleon-nucleon
collisions in a pA collision. It is determined by Glauber theory, which can be subjected
to independent phenomenological tests. The lower left plot of Fig.~\ref{fig1} shows the
nuclear modification factor $R_{d\, Au}^{\pi^0}(p_T, y)$ for the production of neutral pions
in $\sqrt{s_{\rm NN}} = 200$ GeV deuteron-gold collisions at RHIC, calculated within the
factorized ansatz (\ref{eq1}) at leading order (LO). Results shown in Fig.~\ref{fig1} are also 
consistent with the NLO-calculation of $R_{d\, Au}^{\pi^0}(p_T, y)$ in \cite{Eskola:2009uj}.
 All our calculations use LO PDFs from
CTEQ6L~\cite{Pumplin:2005rh} with nuclear modifications EPS09LO~\cite{Eskola:2009uj} and the KKP fragmentation functions~\cite{Kniehl:2000fe}. We have checked our conclusions for another
set of fragmentation functions~\cite{deFlorian:2007aj} (data not shown). 

The $p_T$-dependence of the nuclear modification factor traces the $x$-dependence
of nPDFs.  The precise kinematic connection between the momentum fractions $x_1$, $x_2$
 and the measured hadronic momentum $p_T$ is complicated by the convolution of the
 distributions in (\ref{eq1}). Qualitatively, at fixed rapidity $y$ of the 
 produced hadron, increasing $p_T$ tests larger values of $x_1$, $x_2$.
  Inspection of the nuclear modification factor 
 in the lower left panel of Fig.~\ref{fig1} reveals that 
 the enhancement of $R_{d\, Au}^{\pi^0}(p_T, y)$ in the region around $p_T \simeq 4$ GeV
 at mid-rapidity tests momentum fractions in the anti-shadowing region. 
 The RHIC data~\cite{Adler:2006wg} in Fig.~\ref{fig1} have 
 been used in constraining the nPDF analysis EPS09 \cite{Eskola:2009uj}
 but they were not employed in a closely related nPDF fit~\cite{Eskola:1998df}, which
 provides an equally satisfactory description of these RHIC data.  
Therefore, the agreement of data and calculation  in 
 Fig.~\ref{fig1} is in support of collinear factorization. 
 
 However, qualitatively different explanations of the $R_{d\, Au}^{\pi^0}(p_T, y)$
 measured at RHIC are conceivable. The above calculation accounted for 
 $R_{d\, Au}^{\pi^0}(p_T, y=0)$ in terms of a 
nuclear modification of the {\it longitudinal} parton momentum distribution, only. 
Alternatively, it has been suggested (see e.g.~\cite{Zhang:2001ce}) 
that the characteristic enhancement of $R_{pA}^{h}(p_T, y)$ in the
$p_T$-range of a few GeV (typically referred to as Cronin effect~\cite{Cronin:zm})
can be understood in terms of {\it transverse} parton momentum 
broadening induced by multiple scattering.
 Transverse nuclear broadening is the prototype of a
generic nuclear modification, for which we do not know whether and how it could be  
absorbed in collinear, process-independent nPDFs. 
How can one test whether the physics underlying 
$R_{p\, A}^{h}(p_T, y)$ can be attributed to a nuclear modification of longitudinal parton 
momentum distributions and thus can indeed provide reliable quantitative constraints on 
nPDFs? To address this question, we have calculated  $R_{p\, Pb}^{\pi^0}(p_T, y)$ for the 
production of neutral pions in proton-lead collisions at the LHC, see the right hand 
side of Fig.~\ref{fig1}. 

The LHC can collide protons and Pb ions with a maximum center of mass energy of
$\sqrt{s_{\rm NN}} = 8.8\, {\rm TeV}$. While pPb is not yet part of the 
initial LHC program, there are estimates~\cite{Accardi:2004be}  that without major upgrades a 
luminosity of ${\cal L}_{p\, Pb} = 10^{29}\, {\rm cm}^{-2}\, {\rm s}^{-1}$ could be achieved. 
With these assumptions, we find that running the LHC for one month would allow one to
map out the single inclusive $\pi^0$-spectrum up to transverse momenta well above
$p_T \simeq 50$ GeV (see Fig.~\ref{fig1}). 

Remarkably, if the entire nuclear effect in pPb collisions can be factorized
into nPDFs, then the shape of the nuclear modification factor measured at the
LHC will be qualitatively different from that observed at RHIC. This is so because at 
more than 40 times higher center of mass energy, final state hadrons at the same transverse 
momentum test O(40) times smaller momentum fractions $x_i$. 
As a consequence, $R_{p\, Pb}^{\pi^0}(p_T, y=0)$ at the LHC will be dominated by the shadowing regime, and thus show a suppression, for  $p_T \lesssim 10\div20$ GeV, whereas RHIC data show a clear enhancement in this region.
Further,  LHC data will show a nuclear enhancement
in the anti-shadowing dominated range of $p_T \gtrsim 10\div20$ GeV, whereas the RHIC nuclear
modification factor starts being dominated by $x$-values in the EMC-regime.

We emphasize that a shift of the maximum of 
$R_{p\, Pb}^{\pi^0}(p_T, y=0)$ to values of $p_T > 50$ GeV at the LHC is a natural
consequence of nuclear modifications in {\it longitudinal} parton momentum distributions,
as encoded e.g. in EPS09. In contrast,  no mechanism is known which could account for such 
a large $p_T$-shift in terms  of {\it transverse} parton momentum broadening; 
the  $\sqrt{s}$-dependence of transverse momentum broadening is much
milder. The inverse is equally true: a mild shift of the maximum of 
$R_{p\, Pb}^{\pi^0}(p_T, y=0)\vert_{\sqrt{s_{\rm NN}}= 8.8\, {\rm TeV}}$ to values of 
$p_T \leq 10$ GeV could not be accommodated naturally in a collinear factorized approach, 
since it would imply a nuclear enhancement of some PDFs below $x \simeq 0.01$,
which is inconsistent with the position of the anti-shadowing region.
However, such a mild shift would be
a natural consequence of transverse momentum broadening. 

We also emphasize that at the LHC, a collinearly factorized approach results 
typically in a mild enhancement of $R_{p\, Pb}^{\pi^0}(p_T, y=0)$ above unity for a wide 
transverse momentum range $p_T > 10$. In this kinematic range, suppression factors of 
order $2$ are inconsistent with all existing nPDFs. In contrast, models based on
non-linear small-x evolution (see e.g. Ref.~\cite{Albacete:2010bs}) 
arrive naturally at such large suppression factors. 

The two examples mentioned above illustrate how the much wider kinematic range 
accessible in pPb collisions at the LHC allows one to discriminate decisively
between qualitatively different models of nuclear modification.

%
\begin{figure}
\centering
\includegraphics[width=1.0\linewidth]{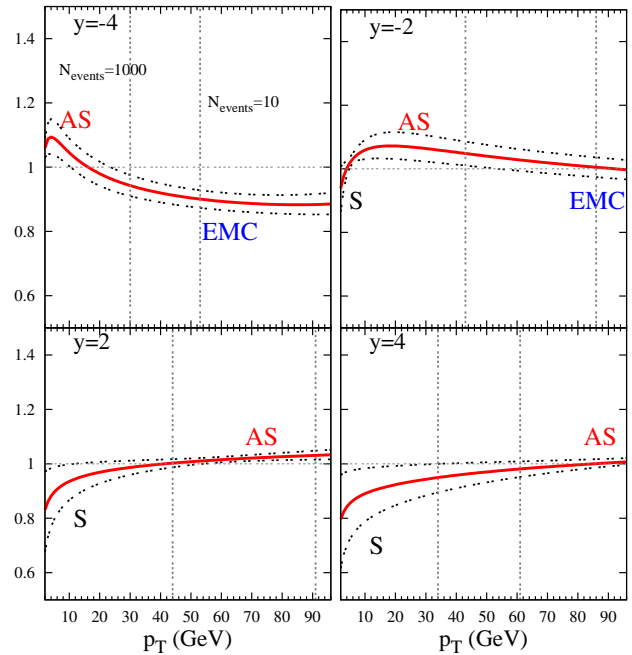}
\vskip -1.2 cm
\caption{\label{fig2} Rapidity dependence of $R_{pPb}^{\pi^0}$ (\ref{eq3}) 
for $\sqrt{s_{\rm NN}}= 8.8$ TeV pPb (LHC). The different plots scan the dependence
from $y = -4$ (close to Pb projectile rapidity) up to $y = 4$ (close to proton projectile rapidity).
Labels indicate whether the nuclear modification originates mainly from the shadowing (S),
anti-shadowing (AS) or EMC regime. Vertical lines illustrate the rapidity-dependent $p_T$
range, which can be accessed experimentally with more than $N_{\rm events} = 1000$ (= 10)
per GeV-bin within one month of running at nominal luminosity. 
}
\vskip -0.5 cm
\end{figure}

The dynamical explanation of the nuclear modification factor 
$R_{p\, Pb}^{h}$ in terms of process-independent collinearly factorized nPDFs 
implies that, as a function of $\sqrt{s_{\rm NN}}$ and rapidity $y$, 
the same nuclear effect manifests itself in very different kinematic ranges of pA collision data.
As seen in Fig.~\ref{fig2}, the rapidity 
dependence of $R_{p\, Pb}^{h}$ allows one to scan the main qualitatively different ranges
of standard nPDFs in an unprecedented way. At backward proton projectile rapidity ($y = -4$),
where relatively large nuclear momentum fractions $x$ are required for hadron production,
the nuclear effects in Fig.~\ref{fig2} are seen to be dominated by the anti-shadowing regime
at low transverse momentum $p_T < 20$ GeV and by the EMC suppression
at higher $p_T$. As one moves to larger rapidity, where smaller nuclear momentum fractions
dominate hadron production, the anti-shadowing regime contributes up to increasingly high-$p_T$,
and opens up a wide window of transverse momentum, in which the shadowing region of
nPDFs can be tested experimentally. 

Within the collinearly factorized approach, one expects non-perturbative
corrections to the ansatz (\ref{eq1}). However, these corrections die out as inverse powers of the 
resolution scale. In contrast, while nuclear effects in nPDFs also depend on the resolution scale, 
their dependence is only logarithmic, so that sizeable nuclear effects are expected to
persist at perturbatively large $p_T$-scales. Therefore,
concise tests of the collinearly factorized approach require particle
production processes at sufficiently large `perturbative' momentum transfers. In pPb collisions 
at the LHC, the experimental access to a wide, nominally perturbative $p_T$-range 
($p_T > 10$ GeV, say) is thus a qualitative advantage. In particular, the forward rapidity 
dependence of  RHIC data \cite{Arsene:2004ux,Back:2004bq} on 
$R_{d\, Au}^h$ does not yet provide 
a decisive test for the collinearly factorized approach, since they test relatively low resolution 
scales, where large corrections to (\ref{eq1}) could be expected even {\it within} the framework of a collinear factorized approach. For this reason these data have not been included in recent  
nPDF analyses~\cite{Eskola:2009uj}. In contrast, {\it within} the framework of a collinearly 
factorized approach, one does not know of sizeable corrections to (\ref{eq1}) in the
range $20 < p_T < 40$ GeV, which will be uniquely accessible at LHC and where
characteristic rapidity-dependent features are seen in Fig.~\ref{fig2}.
%
\begin{figure}
\centering
\includegraphics[width=1.0\linewidth]{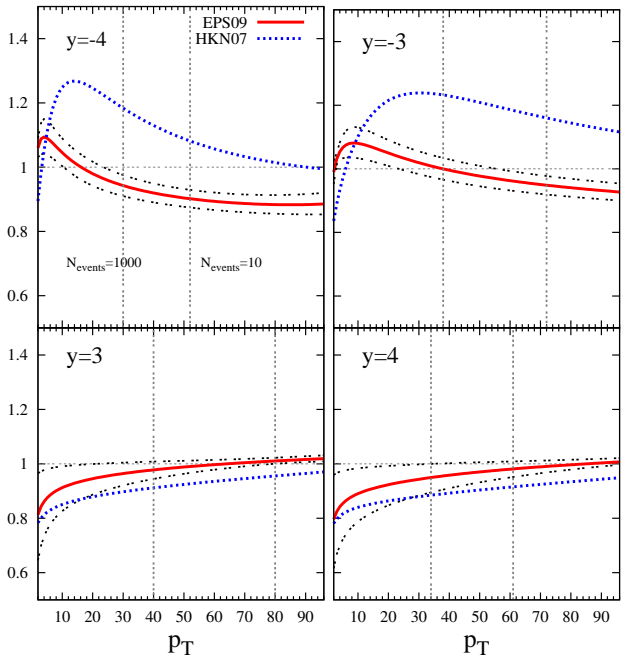}
\vskip -1.3 cm
\caption{\label{fig3}  $R_{pPb}^{\pi^0}$ (\ref{eq3}) 
for $\sqrt{s_{\rm NN}}= 8.8$ TeV pPb (LHC), from two different
sets of nPDFs. 
}
\vskip -0.5 cm
\end{figure}

So far, we have emphasized that well beyond quantitative improvements, pPb
collisions at the LHC have the potential to submit the very assumption of collinear 
factorization to decisive tests. In particular, a strong suppression of 
$R_{p\, Pb}^h(p_T,y=0)$ at high $p_T>10$ GeV, or the persistence of the 
maximum of $R_{p\, Pb}^h(p_T,y=0)$ at $p_T<10$ GeV
is inconsistent with {\it all} current nPDFs and it tests an $x$-range for which 
existing data provide constraints. Therefore, if observed, such features would 
shed significant doubt on the use of the factorized ansatz (\ref{eq1}) for calculating nuclear 
effects, while they could be accounted for naturally in
the context of qualitatively different dynamical explanations, mentioned above.

Despite these perspectives for qualitative tests of collinear factorization, we caution
 that current global analyses of nPDFs come with significant uncertainties. 
While not all conceivable data on $R_{p\, Pb}^h(p_T,y)$ at the LHC 
can be accommodated within a collinearly 
factorized approach,  a significant spread could.
To illustrate this, we have compared in Fig.~\ref{fig3} the nuclear modification factor
for two nPDF sets, which are known to show marked differences. In particular, in contrast
to EPS09, the gluon distribution of HKN07~\cite{Hirai:2007sx} does not show an 
anti-shadowing peak but turns for $x > 0.2$ from suppression to strong 
enhancement  at initial scale $Q^2 = 1\, {\rm GeV}^2$. Inspection of Fig.~\ref{fig3}
reveals that for HKN07, the size and position of the maximum of  $R_{p\, Pb}^h(p_T,y)$
at negative $y$ arises from an interplay between the nuclear enhancement of the 
gluon PDF (which increases with $x$ and hence with $p_T$) and the relative 
contribution of the gluon versus the quark distribution to $R_{p\, Pb}^h(p_T,y)$ 
(which decreases with $p_T$).  Fig.~\ref{fig3} thus illustrates that
within the validity of a collinearly factorized approach, LHC data can 
resolve the qualitative differences between existing nPDF analyses and can
improve significantly and within a nominally perturbative regime on our knowledge of nuclear 
gluon distribution functions. Data on other single inclusive particle spectra and jets in
pPb at the LHC can further constrain global nPDF analysis, thereby testing the concept
of collinear factorization of nuclear effects {\it and} improving our knowledge of nPDFs
as long as this test is passed.

\noindent 
{\bf Acknowledgments.}
We thank N. Armesto, D. d'Enterria, K. Eskola, H. Paukkunen and C. Salgado for helpful
discussions. 
We acknowledge support from MICINN (Spain) under project FPA2008-01177 and FPU grant; Xunta de Galicia (Conselleria de Educacion) and through grant PGIDIT07PXIB206126PR, the Spanish Consolider-
 Ingenio 2010 Programme CPAN (CSD2007-00042) and Marie Curie MEST-CT-2005-020238-EUROTHEPHY (PQA); and Funda\c c\~ao para a Ci\^encia e a Tecnologia (Portugal) under project CERN/FP/83593/2008 (JGM).




\end{document}